\documentclass{aastex}          
\usepackage{spr-astr-addons}    

\begin{document}
%
\title{The duration distribution of Swift Gamma-Ray Bursts}

\shorttitle{<Duration distribution of Swift GRBs>}
\shortauthors{<Horv\'ath and T\'oth>}

\author{Author\altaffilmark{}} 

\author{I. Horv\'ath \altaffilmark{1}} 
\and 
\author{B. G. T\'oth\altaffilmark{1}}


\altaffiltext{1}{National University of Public Service, Budapest, Hungary
              \email{horvath.istvan@uni-nke.hu}}

\begin{abstract}
Decades ago two classes of gamma-ray bursts were identified 
and delineated as having durations shorter and longer than about 2 s. 
Subsequently indications also supported the existence of a third class. 
Using maximum likelihood estimation we analyze the duration 
distribution of 888 Swift BAT bursts observed before October 2015. 
Fitting three log-normal functions to the duration distribution of the 
bursts provides a better fit than two log-normal distributions, 
with 99.9999\% significance. 
Similarly to earlier results, we found that a fourth component is not needed.
The relative frequencies of the distribution of the groups 
are 8\% for short, 35\% for intermediate and 57\% for long bursts
which correspond to our previous results. 
We analyse the redshift distribution for the 269 GRBs
of the 888 GRBs with known redshift.
We find no evidence for the previously suggested difference
between the long and intermediate GRBs' redshift distribution.
The observed redshift distribution of the 20 short GRBs differs with high
significance from the distributions of the other groups.
\end{abstract}


%
\section{INTRODUCTION}
Decades ago 
\citep{maz81} and \citep{nor84} suggested that there is a
separation in the duration distribution of gamma-ray bursts (GRBs). 
Today it is
widely accepted that the physics of the short and long GRBs are different, and
these two kinds of GRBs are different phenomena
\citep{nor01,bal03,fox05,zha09,lu10}.

In the Third BATSE Catalog \citep{m6} --- using uni- and
multi-variate analyses --- \citep{ho98} and \citep{muk98} found a third
type of GRBs.  Later several papers
\citep{hak00,bala,rm02,ho02,hak03,bor04,ho06,ch07} confirmed the
existence of this third ("intermediate" in duration) group in
the same database. 
In the Swift data
the intermediate class has also been found \citep{ho08,hu09}. 
There are also more recent works \citep{ho09,bala11,up11,lu14} in this field.

In the Swift database, 
the measured redshift distribution
for the two groups are also different, for short burst the
median is 0.4 \citep{os08} and for the long ones it is 2.4
\citep{bag06}.

The paper is organized as follows. Section 2
briefly summarizes the method and fits,
Section 3 contains the calculations of the fits and
their results, Section 4 contains the comparison of the 
redshift distributions of the different classes 
and Section 5 summarizes the
conclusions of this paper.

\section{THE METHOD}

On the Swift web page\footnote{http://swift.gsfc.nasa.gov/archive/grb table} 
there are 997 GRBs; 957 of these bursts were catalogued prior to October 2015. 
Multi-variate analyses have demonstrated that flux and fluence values are also 
needed for a robust classification.
There were 30 GRBs without fluence information
and 19 GRBs without peak flux information.
Another 20 GRBs are excluded from this analysis because their 
fluence uncertainties are larger than 50\%.
Figure 1 shows the 
$\log T_{90}$ (time to accumulate the central 90\% 
of the burst fluence) distribution of the remaining 888 GRBs.

The maximum likelihood (ML) method assumes that the probability density function of an
$x$ observable variable is given in the form of $g(x,p_1,...,p_r)$
where $p_1,...,p_r$ are parameters of unknown value. Having  $ N $
observations on $x$, one can define the likelihood function in the
following form:

\begin{equation}
l = \prod_{i=1}^{N}  g(x_i,p_1,...,p_r) ,\
\end{equation}

\noindent or in logarithmic form (the logarithmic form is more
convenient for calculations):

\begin{equation}
L= \log l = \sum_{i=1}^{N} \log  \left( g (x_i,p_1,...,p_r) \right).
\end{equation}

The ML procedure maximizes $L$ according to
$p_1,...,p_r$.  Since the logarithmic function is monotonic, the
logarithm reaches its maximum at the same
$p_1,...,p_r$ parameter set where $l$ does.
The confidence region of the estimated parameters is given by the
following formula, where $L_{max}$ is the maximum value of the
likelihood function and $L_0$ is the
likelihood function  at the true value of the
parameters \citep{KS76}:

\begin{equation}
2 ( L_{max} -  L_0) \approx \chi^2_r, \label{eq:chi}.
\end{equation}

\section{LOG-NORMAL FITS OF THE DURATION DISTRIBUTION }

\begin{table}
\caption[]{The best parameters for the two log-normal fit of
the GRB duration distribution. } \label{2g}
$$
         \begin{array}{cccc}
            \hline
            \noalign{\smallskip}
     & center (log T_{90})  &  \sigma  (log T_{90}) &  w \\
            \noalign{\smallskip}
            \hline
            \noalign{\smallskip}
        short      &   0.386   &  0.484   &   188.1   \\
        long      &     1.700   &  0.480   &  699.9    \\
            \noalign{\smallskip}
            \hline
         \end{array}
     $$
\end{table}

\begin{figure}
\centering
\resizebox{\hsize}{!}
{\includegraphics[angle=0,width=7cm]{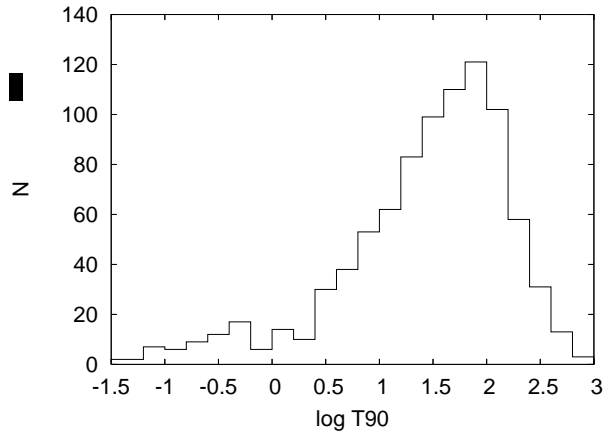}}
\caption{The duration distribution of Swift BAT bursts.  }
\end{figure}

Similar to our previous work \citep{ho02,ho08}, we 
fit the $T_{90}$  distribution using
ML with a superposition of $k$ log-normal components, each
of them having two unknown parameters to be fitted with $N=888$
measured points. 
The choice to use log-normal functions to fit the duration 
distribution is based  on the results of \citep{bal03,sn15,zit15}.
Our goal is to find the minimum value
of $k$ suitable to fit the observed distribution. Assuming a
weighted  superposition of $k$ log-normal distributions, one has to
maximize the following likelihood function:

\begin{equation}
L_k = \sum_{i=1}^{N} \log  \left(\sum_{m=1}^k   w_mf_m(x_i,\log
T_m,\sigma_m ) \right)
\end{equation}

\noindent where $w_m$ is a weight and $f_m$ is a log-normal function with
$\log T_m$ mean and $\sigma_m $ standard deviation having the form
of

\begin{figure}
\centering
\resizebox{\hsize}{!}
{\includegraphics[angle=0,width=7cm]{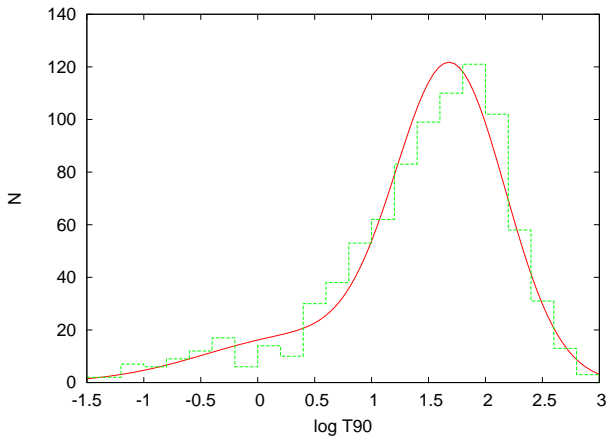}}
\caption{Fit with two log-normal component for the duration distribution of BAT bursts.  }
\label{fit3}
\end{figure}

\begin{equation}
$$f_m = \frac{1}{ \sigma_m  \sqrt{2 \pi  }}
\exp\left( - \frac{(x-\log T_m)^2}{2\sigma_m^2} \right)  $$
\label{fk}
\end{equation}

\noindent and due to a normalization condition

\begin{equation}\label{wight}
    \sum_{m=1}^k w_m= N \, .
\end{equation}

 \noindent We used a simple
C++ code to find the maximum of $L_k$. Assuming only one log-normal
component, the fit gives $L_{1max}=4978.88$ but in the case of
$k$ = 2 one gets $L_{2max}=5082.246$ with the parameters given in
Table~1. The solution is displayed in Fig.~2.

Based on Eq. (\ref{eq:chi}) we can infer whether the addition of a
further log-normal component is necessary to significantly improve
the fit. We make the null hypothesis that we have already reached 
the the true value of $k$. Adding a new component, i.e. moving
from $k$ to $k+1$, the ML solution of $L_{kmax}$ has changed to
$L_{(k+1)max}$, but $L_0$ remained the same. In the meantime we
increased the number of parameters with 3 ($w_{k+1}$, $logT_{k+1}$
and $\sigma_{(k+1)})$. Applying Eq. (\ref{eq:chi}) on both
$L_{kmax}$ and $L_{(k+1)max}$ we get after subtraction

\begin{equation}\label{kk1}
2 ( L_{(k+1)max} -  L_{kmax}) \approx \chi^2_3 \, .
\end{equation}

\noindent For $k=1$, $L_{2max}$ is greater than $L_{1max}$
by more than 100, which gives for $\chi^2_3$ an
extremely low probability. It means that the fit 
with two log-normal distributions is really a better approximation for
the duration distribution of GRBs than the fit with one.

\begin{figure}
\centering
\resizebox{\hsize}{!}
{\includegraphics[angle=0,width=7cm]{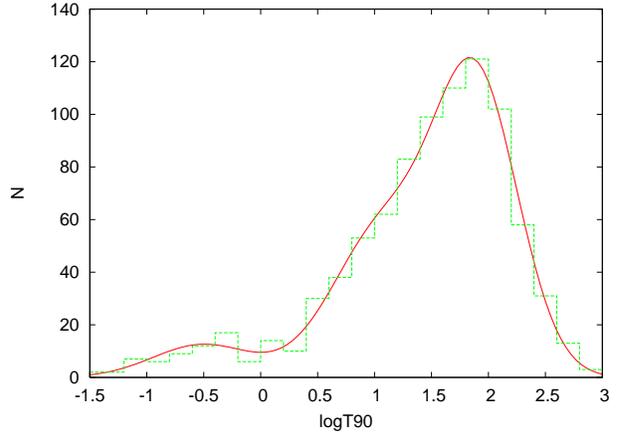}}
\caption{Fit with three log-normal components for the duration distribution of BAT bursts.  }
\label{fit3}
\end{figure}

Thirdly, a
three-log-normal fit is made combining three $f_k$ functions with
eight parameters (three means, three standard deviations and two independent
weights).  For the best fitted parameters see Table 2. The highest value of the
logarithm of the likelihood ($L_{3max}$) is 5098.361.  
For two log-normal functions the maximum is 5082.246. The maximum has thus improved 
by 16. Twice this value is 32, which gives us a probability of 0.00006\% for 
the difference between $L_{2max}$ and $L_{3max}$ being only by chance. Therefore, 
there is a high probability that a third log-normal distribution is needed. 
In other words, the three-log-normal fit
(see Fig. 3) is better and there is a 0.0000006
probability that it was caused only by statistical fluctuation.

\begin{table}
\caption[]{The best parameters for the three log-normal fit of the
GRB duration distribution.}
         \label{3g}
     $$
         \begin{array}{cccc}
            \hline
            \noalign{\smallskip}
     & center (log T_{90})  & \sigma  (log T_{90}) &  w \\
            \noalign{\smallskip}
            \hline
            \noalign{\smallskip}
          short   &  -0.508  &  0.439   &   68.9   \\
           long   &   1.897   & 0.367   &  506.3   \\
    intermediate  &   1.076   & 0.448   &  312.8    \\
            \noalign{\smallskip}
            \hline
         \end{array}
     $$

\end{table}

In one of our previous papers \citep{ho08}, we published a similar
analysis on 222 GRBs of the First BAT Catalog.
One should compare these results with the results published in that paper. 
The centers of the distributions change by only a very small amount. 
In the current analysis, 
the center of the distribution of the short bursts
is at -0.508 (0.311 s) which was previously at -0.473 (0.336 s). For the intermediate
ones, the center is at 1.076 which was at 1.107 and for
the long bursts at 1.897, which was at 1.903 in our previous analysis.
The relative frequencies of the distribution of the groups now
are 8\% for short, 35\% for intermediate and 57\% for long bursts (see Table 2).
Using a sample four times smaller in 2008 \citep{ho08},
the relative frequencies of the distribution of the groups 
were 7\% for short, 35\% for intermediate and 58\% for long bursts. 
Therefore, both analyses give us very similar results.
This does not mean that the three-Gaussian is the
best approximation for the duration distribution of the GRBs,
but strongly suggests that the smaller and the larger sample
duration distributions are almost the same.
Neither the nature of the GRBs nor the Swift
detectors changed during the years.

One should also calculate the likelihood for
four log-normal functions. The best logarithm of the ML is 5098.990.
It is larger with 0.63 than it was for three log-normal functions.
This gives us a low significance (26\%),
therefore the fourth component is not needed. In Table 3
we summarize the improvement of the likelihood and
the corresponding significances.

   \begin{table}
 \caption[]{The improvement of the likelihood and
the significances.}
         \label{4g}
     $$
         \begin{array}{cccc}
            \hline
            \noalign{\smallskip}
   i  & L_{imax}  & L_{imax} - L_{(i-1)max} &  significance \\
            \noalign{\smallskip}
            \hline
            \noalign{\smallskip}
          2   &  5082.246  &     &     \\
           3   &  5098.361   & 16.1   &  0.999999    \\
    4  &   5098.99   & 0.63   & 0.26     \\
            \noalign{\smallskip}
            \hline
         \end{array}
     $$

   \end{table}

\section{THE REDSHIFT DISTRIBUTIONS }

Among the 888 GRBs there are 269 GRBs which
have redshift information.
Since the duration distribution of the three groups
overlap one cannot be sure to which group does a
specific burst belong.
However, 2.5 s (0.4 in logarithmic scale) and 31.6 s (1.5 in logarithmic scale)
seems to be an approximate border between the short,
intermediate and long GBRs.
Using these borders, there are 20 short, 79 intermediate
and 170 long GRBs.
We cut the biggest population, the long burst group,
into two parts: shorter than 100 s (91 GRBs) and longer than 100 s (79 GRBs).
Figure 4 shows the cumulative distribution of the
short, intermediate, long1 and long2 bursts.
One can use the Kolmogorov-Smirnov (KS) test to
compare the distributions.
The redshift distribution of the intermediate, the long1 and the long2 groups
do not differ from each other.
The probabilities can be seen in Table 4.
However, the observed redshift distribution of the short bursts
differs from the other three distributions
with high significance (more than 99.9 \%).

\begin{figure}
\centering
\resizebox{\hsize}{!}
{\includegraphics[angle=0,width=7cm]{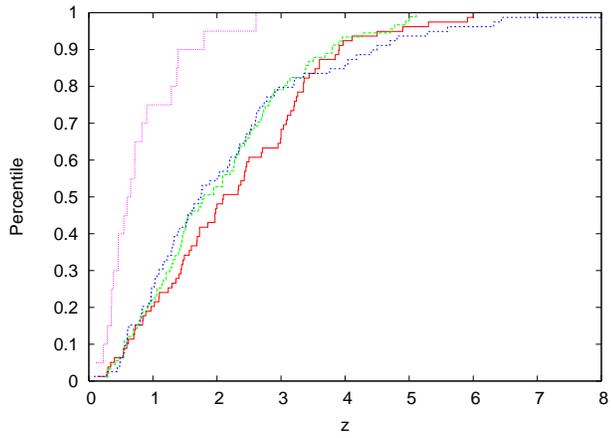}}
\caption{The cumulative redshift distribution of the
short (pink), intermediate (blue), long1 (green)
and long2 (red) bursts.  }
\label{fit3}
\end{figure}
 
This latter result is well-known in the literature.
For the intermediate bursts, there was a
suggestion with a very low significance
that the redshift distribution of these bursts
are different from the redshift
distribution of the long GRBs.
Based on the results discussed above,
we are not able to confirm this suggestion.
The redshift distributions of the long and the
intermediate GRBs seem to be very similar.
 
   \begin{table}
 \caption[]{The KS probabilities assuming the same redshift
 distribution for the classes.}
         \label{4g}
     $$
         \begin{array}{cccc}
            \hline
            \noalign{\smallskip}
   p(KS)  & intermediate  & long1 &  long2 \\
            \noalign{\smallskip}
            \hline
            \noalign{\smallskip}
          short   &  0.000  &  0.000   &  0.000   \\
         intermediate   &    & 0.90   &  0.22    \\
    long1  &     &    & 0.21     \\
            \noalign{\smallskip}
            \hline
         \end{array}
     $$
 
   \end{table}
 
\section{CONCLUSIONS }
 
We presented that fitting the duration distribution of 888 Swift BAT GRBs
with three log-normal functions is better than the
fit using only two. Though this may be the result of statistical fluctuations
and maybe there are only two types of GRBs, the
probability that the third component is not needed is only 0.00006\%.
One can compare the parameters of the burst groups with previous results.
In \citep{ho08} the relative frequencies
were 7\% for short, 35\% for intermediate and 58\% for long bursts.
Now our results show
8\% for short, 35\% for intermediate and 57\% for long ones.
The center of the groups are also nearly the same.
 
We have shown with very high significance that
the redshift distribution of the short bursts is
different from the redshift distributions of the other (longer) GRBs.
However, the redshift distribution of the intermediate GRBs seems to
be similar to the redshift distribution of the long GRBs.

%
\acknowledgments
This research was supported by OTKA grant NN111016.


%

%

\end{document}